\documentclass[]{kluwer}
\usepackage[dvips]{graphicx}
\begin{document}
\begin{article}
\begin{opening}
\title{Evolution and Fragmentation of Wide-Angle Wind Driven Molecular
Outflows}
\subtitle{}

\author{Andrew~\surname{Cunningham}}
\author{Adam~\surname{Frank}}
\author{Peggy~\surname{Varni\`ere}}
\author{Alexei~\surname{Poludnenko}}
\institute{University of Rochester}
\author{Sorin~\surname{Mitran}}
\institute{University of North Carolina, Chapel Hill}
\author{Lee~\surname{Hartmann}}
\institute{Harvard-Smithsonian Center for Astrophysics}
\begin{abstract}
We present two dimensional cylindrically symmetric hydrodynamic
simulations and synthetic emission maps of a stellar wind
propagating into an infalling, rotating environment.  The
resulting outflow morphology, collimation and stability observed
in these simulations have relevance to the study of young stellar
objects, Herbig-Haro jets and molecular outflows.  Our code
follows hydrogen gas with molecular, atomic and ionic components
tracking the associated time dependent molecular chemistry and
ionization dynamics with radiative cooling appropriate for a dense
molecular gas.  We present tests of the code as well as new
simulations which indicate the presence of instabilities in the
wind-blown bubble's swept-up shell.
\end{abstract}

\keywords{Protostellar Outflow, HH Object, Molecular Outflow,
Wide-Angle Wind, Fragmentation}
\end{opening}

\section{Introduction}

Bipolar Jets and wide angle molecular outflows are recognized as a
ubiquitous phenomena associated with star formation.  It is
expected that most if not all low mass stars produce such outflows
during their formation through the gravitational collapse of gas
from the parent molecular cloud.  A molecular outflow is formed
when molecular gas is displaced from the cavity evacuated by a
fast stellar wind.  This results in the formation of irregular
lobes and thin shells of swept up shocked molecular gas along the
walls of the cavity.  The strong radiative energy loss from the
shock heated molecular gas can result in the onset of several
instabilities in the molecular outflow \cite{v}, \cite{vr}.  We
present some preliminary results of our work employing
multidimensional numerical models including molecular chemistry
and associated radiative losses to explore the fragmentation and
stability properties of these outflows.

Simulations of protostellar outflows in the presence of a
collapsing molecular core have been carried out using the
AstroBEAR adaptive mesh refinement muti-physics code \cite{pol},
\cite{var}.  The AstroBEAR code employs an exact hydrodynamic
Riemann solver and a conservative integration scheme in an
Eulerian frame of reference to advance the solution of the source
free Euler equations.  The geometric and microphysical source
terms are split from the hydrodynamic integration using an
implicit fourth-order Rosenbrock source term integration scheme
for stiff ODE's.  The use of adaptive mesh code has been essential
to achieve the necessary resolution in the neighborhood of thin
shock bounded high density slabs that are prevalent in these
simulations.

\section{Isothermal X-wind Model Simulation}

\begin{figure}[!h]
\begin{center}
\includegraphics[width=0.7\textwidth]{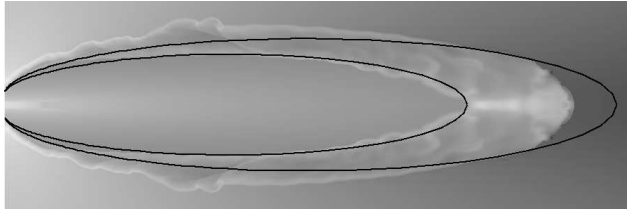}
\caption{X-wind outflow at t = 304~yr.  The outer curve represents
the analytically predicted shell morphology at this time.  The
inner curve follows the shape of the analytical model at an
earlier time and whose morphology matches the shape of the inner
shock in our simulations. }\label{leetest}
\end{center}
\end{figure}

We first compare our code with results of previous calculations.
In figure \ref{leetest} we show a simulation of an isothermal
outflow with conditions similar to the X-wind model of
\inlinecite{sno} driving into a toroidal ambient medium. This
simulation uses a velocity and density pattern similar to the
X-wind but no magnetic field is included. We compare figure
\ref{leetest} with the results of the same calculation performed
by \inlinecite{lso} at 1/4 the resolution of the current work.  Of
particular interest is the extent to which the late-time flow
morphology depicted in figure \ref{leetest} agrees with the
analytically predicted result of \inlinecite{lso} delineated by
the black curve in the figures as a verification of our code in
these regimes.

The higher resolution in our simulation allows us to track the
internal dynamics of the swept-up shell.  Our simulations reveals
fine scale features generated as the material flows along the
shell walls something not possible with the lower resolution
study. Note that the analytic shell morphology model assumes a
thin shell. The violation of this conditions is likely responsible
for the deviation from the analytically predicted morphology given
by the outer curve.  The general form of the outflow morphology
is, however, well approximated by the analytic model.

\section{Protostellar Wind-Infall Model with Non-Equilibrium
Ionization, $H_2$ Chemistry and Cooling}

\begin{figure}[!h]
\begin{center}
\includegraphics[width=0.7\textwidth]{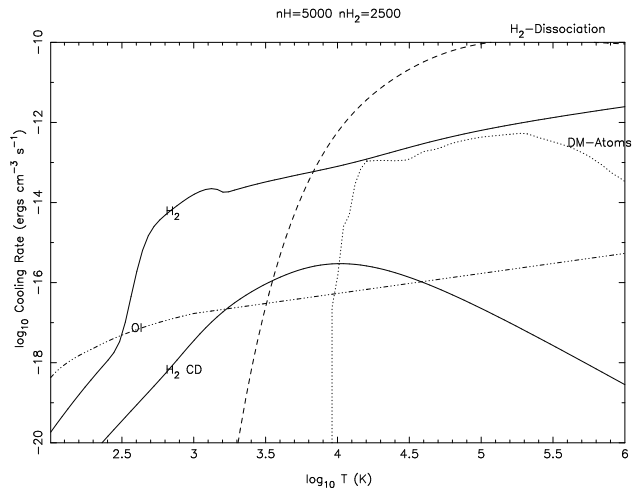}
\caption{Cooling rates for typical ISM abundances,
$n_{H_{2}}=2500$,$n_{HI}=5000$. $H_2$ is the molecular hydrogen cooling
function for a tenuous gas, $H_2$ CD is the molecular cooling
function valid for $n >> n_{critical}$, $H2-Dissociation$ is the thermal
energy loss due to the dissociation of $H_2$ molecules, $OI$ is the
singly ionized oxygen line cooling and $DM-atoms$ is the atomic line
and recombination cooling function of Dalgarno \& McCray.
}\label{cooling}
\end{center}
\end{figure}

\begin{figure}[!h]
\begin{center}
\includegraphics[width=0.7\textwidth]{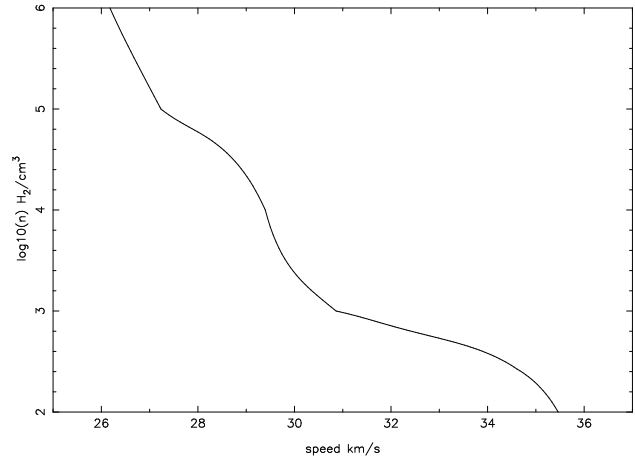}
\caption{Preshock density vs. shock speed for a steady shock resulting
in 90\% downstream $H_2$ dissociation.}\label{disspeed}
\end{center}
\end{figure}

We have also performed simulations including the effects of
non-equilibrium HI ionization \cite{ar}, \cite{hm}, \cite{m}, H2
dissociation and cooling \cite{d}, \cite{ls}, \cite{lrrw},
\cite{mm}, \cite{mk}.  OI line cooling, a dominant cooling agent
at temperatures below 1000~K has also been included \cite{lr}.
Atomic line and recombination cooling has been included using the
cooling rates of a coronal gas \inlinecite{dm}.  The volumetric
cooling rates for the case of a partially ionized gas are plotted
in figure \ref{cooling}.  We have also performed tests of our
chemistry routines.  We show one of these tests where we find the
shock speeds which produce $90\%$ molecular dissociation.   Figure
\ref{disspeed}) shows our code produces results consistent with
those of previous authors \cite{smith}, \cite{hm2}.

\begin{figure}[!h]
\begin{center}
\includegraphics[width=0.65\textwidth]{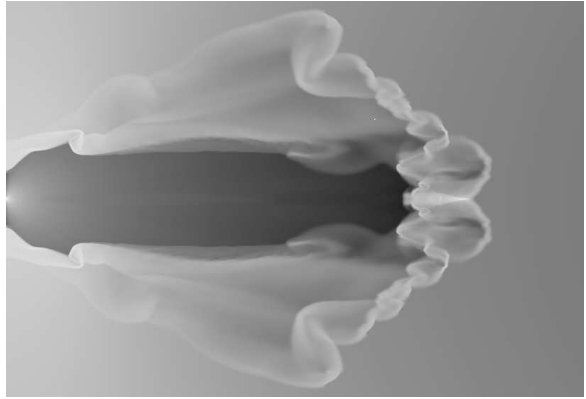}
\caption{Collapsing sheet outflow, low speed case, logarithm of
density at t = 97~yr.  }\label{lospeed}
\end{center}
\end{figure}

Figure \ref{lospeed} shows the interaction of a tenuous, initially
ionized protostellar wind with a cold, slowly rotating molecular
environment in the presence of the central gravitational potential
of .21 $M_{\odot}$ protostar (not resolved in these simulations).
The model for the molecular environment is that of a collapsing of
non-magnetic, self-gravitating sheet of \inlinecite{hcb} as
implemented by \inlinecite{df}.  The initial outflow takes the
form of a sphere ejecting gas at a uniform velocity with an
azimuthal density gradient varying as $cos(\theta)$ and an equator
to pole density contrast of 50.  Thus our inflow condition creates
a wide angle wind in which the momentum input to the environment
by the protostar is aspherical with the bulk of the thrust being
directed along the poles.

The outflow speed in the simulation is maintained at 100~km/s with
the inflow injected through a "wind sphere" of radius of 33~AU, an
outflow rate of $10^{-8} M_{\odot} yr^{-1}$ and an initial infall
rate of $10^{-9} M_{\odot} yr^{-1}$ from the collapsing
environment.  The equatorial outflow ram pressure is overcome by
the gravitational infall of the shocked ambient gas resulting in
appreciable shock focusing of the outflow. This leads to very
efficient collimation of the ejected wind material.

The resolution achieved in these simulations is sufficient to
resolve the swept up shells to 6 to 10 pixels.  This is enough to
track the onset of what appear to be thin shell instabilities. The
shell fragmentation process resulting from such instabilities may
have a significant effect on the efficiency with which the outflow
is able to entrain and disperse its momentum in support of the
collapsing environment and on the morphological signatures of the
outflow.

\begin{figure}[!h]
\begin{center}
\includegraphics[width=0.55\textwidth]{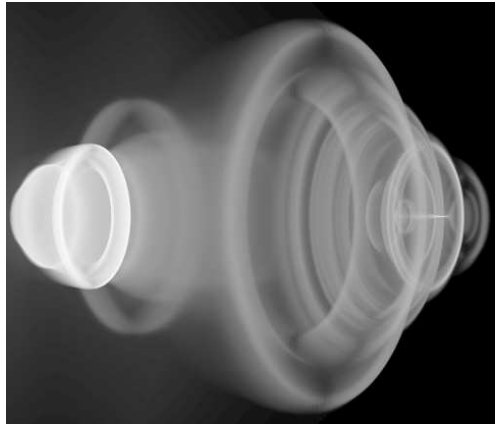}
\caption{Collapsing sheet outflow, low speed case, synthetic $H_2$
emission projection at t = 97~yr.  }\label{synth}
\end{center}
\end{figure}

Figure \ref{synth} shows the 3D $H_2$ emissivity of the outflow
projected onto a plane tilted $30^\circ$ from the symmetry axis of
the outflow.  Note that the brightest emission emanates from the
dissociation region immediately behind the outer bowshock.  Of
course the stability properties of the wind collimation mechanism
as well as the three dimensional nature of the fragmentation of
the thin shell of swept up molecular gas cannot be fully addressed
using the 2D cylindrically symmetric approach taken here. Future
work will focus on the three dimensional stability of the flow
collimation mechanisms as well as on the effect that the outflow
geometry has on the fragmentation of protostellar accretion
shells, morphology and momentum dissipation efficiency.

\section{Conclusions}

We have presented first results of simulations of molecular
outflows using a new Adaptive Mesh Refinement code which tracks
both ionzation and chemistry.  These simulations focus on the
early time evolution of a wide angle wind expanding into
collapsing rotating sheet.  Our simulations are able to marginally
resolve the internal dynamics of the swept-up and with this
resolution we find the leading sections of the outflow lobe to be
unstable to what appear to be Thin Shell modes.  Future work will
focus on exploring the dynamics of the outflows in greater detail
providing links between the early evolution of the outflow and the
late-time large scale appearence.  In particular we are interested
to see if the fragmentation of the shell changes the global
dynamics of the outflow in significant ways by generating a
"clumpy" lobe which expands and sweeps up ambient material.

\end{article}
\end{document}